\documentclass[runningheads]{llncs}
\usepackage{graphicx}
\usepackage[colorlinks]{hyperref}
\usepackage{booktabs}
\usepackage{array}
\usepackage{multirow}
\usepackage{float}
\newcommand{\etal}{\textit{et al.}}
\begin{document}
\title{Extending nn-UNet for brain tumor segmentation}
%
%
\author{Huan Minh Luu\inst{1} \and
Sung-Hong Park\inst{1}}
\authorrunning{Luu et al.}
%
\institute{Magnetic Resonance Imaging Laboratory, Department of Bio and Brain Engineering, Korea Advanced Institute of Science and Technology \\
\email{luuminhhuan@kaist.ac.kr, sunghongpark@kaist.ac.kr}\\
}
\maketitle              
\begin{abstract}
Brain tumor segmentation is essential for the diagnosis and prognosis of patients with gliomas. The brain tumor segmentation challenge has continued to provide a great source of data to develop automatic algorithms to perform the task. This paper describes our contribution to the 2021 competition. We developed our methods based on nn-UNet, the winning entry of last year competition. We experimented with several modifications, including using a larger network, replacing batch normalization with group normalization, and utilizing axial attention in the decoder. Internal 5-fold cross validation as well as online evaluation from the organizers showed the effectiveness of our approach, with minor improvement in quantitative metrics when compared to the baseline. The proposed models won first place in the final ranking on unseen test data. The codes, pretrained weights, and docker image for the winning submission are publicly available.\footnote{\url{https://github.com/rixez/Brats21_KAIST_MRI_Lab} \newline \url{https://hub.docker.com/r/rixez/brats21nnunet}}

\end{abstract}
\section{Introduction}
Brain tumor segmentation from magnetic resonance (MR) images is an essential procedure for brain tumor care, enabling the clinicians to identify the location, extent and types of the tumors. This not only helps with initial diagnosis but also aid with administering and monitoring treatment progress. Given the importance of this task, precise delineation of the tumor and its sub-regions is typically performed manually by experienced neuro-radiologists. This is a tedious and time-consuming process that demands significant effort and expertise, especially when the patient volume is high, the images are multi-parameteric of different contrasts, and the tumors are heterogeneous. The labelling process is also subjected to inter and intra-rater variability \cite{Visser2019}, necessitating a consensus for the labelling and interpretation of the segmentation and add an extra layer of complexity. Automatic or computer-aided segmentation algorithms have the potential to resolve these shortcomings as it can lower the labor-intensiveness of the labelling process as well as being consistent across different cases. However, to develop these algorithms, sufficient annotated high quality data is needed to achieve performance satisfactory for clinical purposes.

The Brain Tumor Segmentation Challenge (BraTS) \cite{Menze2014,Bakas2018,Bakas2017a,Bakas2017b} is an annual international competition that has been carried out since 2012. Participants are provided with an ample dataset of fully-annotated, multi-institutional, multi-parametric MR images (mpMRI) of patients with varying degrees of gliomas. Since its inception, the dataset has grown from only 30 cases in 2012 to 2000 cases in 2021 \cite{Baid2021}. In this year (2021) challenge, the participants can compete in two tasks: the usual task of brain tumor sub-region segmentation from mpMRI and the novel task of prediction of MGMT (0[6]-methylguanine-DNA methyltransferase) promoter methylation status, which is an important biomarker to determine the response of the patients to cancer treatment. We participated only in the segmentation task and this manuscript describes our entry in this part of the competition.

Initial attempts toward automatic segmentation of brain tumors have relied on hand-crafted features engineering with traditional machine learning methods such as Atlas-based \cite{Bauer10}, decision forest\cite{Zikic12,Tustison13}, conditional random field \cite{Wu14}. With the rising popularity of deep learning enabled by the improvement in computational capability of modern graphic processing units (GPU), the efficiency of algorithms, and the availability of training data, conventional machine learning methods have slowly been replaced by deep neural networks in several fields such as computer vision \cite{Alex12,Kaiming15}, natural language processing \cite{Vaswani17}, or computational biology \cite{Jumper21}. In the context of the BraTS competition, deep learning algorithms have been explored for tumor segmentation since 2014 and they are the algorithm of choice for most entries in recent years. In fact, winners of the 4 most recent competitions all employed deep neural networks, testifying the superior performance of this approach when more data is available. We summarized the main takeaways from these winning contributions. Kamnitsas \etal \cite{Kamnitsas17} proposed Ensemble of Multiple Models and Architecture (EMMA), combining the predictions from different 3D convolutional networks (DeepMedic \cite{DeepMedic1,DeepMedic2}, FCN \cite{Long15}, and U-Net \cite{Ronneberger14}). Myronenko \cite{Myronenko18} combined a 3D U-Net with an additional variational decoder branch to provide additional supervision and regularization to the encoder branch. Jiang \etal \cite{Jiang19} trained a two-stage cascaded U-Net, the first stage was trained to produce coarse segmentation mask and the second stage was trained to refine the output of the first stage. Isensee \etal \cite{Fabian20} utilized nnU-Net \cite{Fabiannat}, a self-configuring framework that automatically adapt U-Net to a particular dataset, and showed robust performance with minimal modifications to the conventional 3D U-Net by utilizing some BraTS-specific optimizations.

For our entry to the competition, we extended the nnU-Net framework proposed by last year winner by adding several components. Due to the ease of adapting nnU-Net to new dataset as well as the fully open-source codes and models, nnU-Net serves as an excellent baseline for further experimentation. The modifications that we explored for this year competition are using group normalization instead of batch normalization, using an asymmetrically large encoder for the U-Net, and using axial attention  in the decoder. Experiment results with cross-validation training data and unseen validation data from the leaderboard showed the effects these modifications have on the performance of the networks. 

\section{Methods}
\subsection{Data}
\begin{figure}
    \centering
    \includegraphics[scale=0.5]{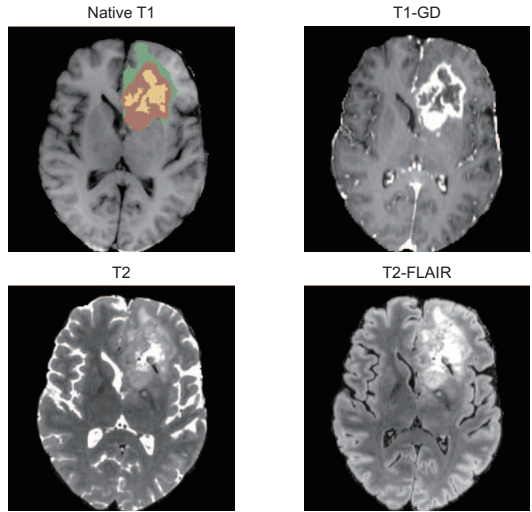}
    \caption{Representative axial slice from one subject (BraTS2021\_00001) showing the four different contrast with overlaid segmentation labels for the 3 tumor sub-regions on the T1 image. Green: peritumoral edematous/invaded tissue, red: necrotic tumor core, yellow: GD-enhancing tumor.}
    \label{fig:1}
\end{figure}

Multi-parametric MRI scans from 2000 patients were used for BraTS2021, 1251 of which were provided with segmentation labels to the participants for developing their algorithms, 219 of which were used for the public leaderboard during the validation phase, and the remaining 530 cases were intended for the private leaderboard and the final ranking of the participants. 4 contrasts are available for the MRI scans: Native T1-weighted image, post-contrast T1-weighted (T1Gd), T2-weighted, and T2 Fluid Attenuated Inversion Recovery (T2-FLAIR). Annotation were manually performed by one to four raters, with final approval from experienced neuro-radiologists. The labels include regions of GD-enhancing tumor (ET), the peritumoral edematous/invaded  tissue  (ED), and the necrotic tumor core (NCR). All MRI scans were preprocessed by co-registration to the same anatomical template, interpolation to isotropic $1mm^{3}$ resolution and skull-stripping. The image sizes of all MRI scans and associated labels are $240 \times 240 \times 155$. Fig. \ref{fig:1} shows a representative slice of the four contrasts with segmentation. Further processing was done on the provided data before inputting into the network. To reduce the computation, the volumes were cropped to non-zero voxels. Since the intensity in MR images is qualitative, the voxels were normalized by their mean and standard deviation.

\subsection{Model}
In this section, the details of the models developed are described, starting from the strong baseline nnU-Net models that won last year competition. Several modifications to this baseline and their rationales are then elaborated. All of these experiments were done with the excellent open-source nnU-Net framework  \footnote{\url{https://github.com/MIC-DKFZ/nnUNet}}.
\subsubsection{Baseline nnU-Net}
nnU-Net by Isensee \etal \cite{Fabian20} was the winning entry for BraTS 2020. At its core is a 3D U-Net that operates on patches of size $128 \times 128 \times 128$. The network has the encoder-decoder structure with skip connections linking the two pathways. The encoder comprises of 5 levels of same-resolution convolutional layers with strided convolution downsampling. The decoder follows the same structure with transpose convolution upsampling and convolution operating on concatenated skip features from the encoder branch at the same level. Leaky ReLU (lReLU) with slope of 0.01 \cite{lrelu} and batch normalization  \cite{bn} was applied after every convolution operations. The mpMRI volumes were concatenated and used as 4-channels input. nnU-Net employs region-based training: instead of predicting three mutually exclusive tumor sub-regions , as in the provided segmentation labels, the network predicts instead the three overlapping regions of enhancing tumor (ET, original region), tumor core or TC (ET + necrotic tumor), and whole tumor or WT (ET + NT + ED). The softmax nonlinearity at the final layer of the network was replaced by a sigmoid  activation, treating each voxels as a multi-class classification problem. As the metrics computed for the public and private leaderboard are based on these regions, this region-based training has been observed to improve the performance. \cite{Fabian20,Zhao20,Myronenko18}. Additional sigmoid outputs were added to every resolution except for the two lowest levels to apply deep supervision and improve gradient propagation to the earlier layers. The number of convolutional filters was initialized at 32 and doubled for every reduction of resolution, up to a maximum of 320.
\subsubsection{Larget network and Group Normalization}
\begin{figure}
    \centering
    \includegraphics[scale=0.35]{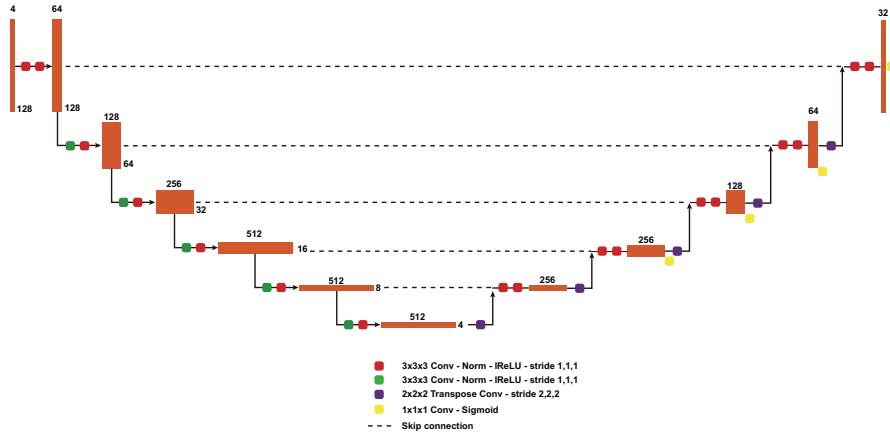}
    \caption{Structure of the larger U-Net with asymmetric scaling on the encoder. The number at the top and bottom indicates the number of channels and the dimension of the features vectors, respectively.}
    \label{fig:2}
\end{figure}

The first modification we made was to increase the size of the network asymmetrically by doubling the number of filters in the encoder while maintaining the same filters in the decoder. This asymmetrically large encoder was utilized by Myronenko \cite{Myronenko18}. As the amount of training data is quadrupled compared to previous year, increasing the capacity of the network will help it able to model the larger data variety. The maximum number of filter was also increased to 512. The structure of the modified network is shown in Fig. \ref{fig:2}. The second modification was to replace all batch normalization by group normalization \cite{GN}. 3D convolutional networks demands a significant amount of GPU memory even with mixed precision training, which limit the batch size that one can use during training. Group normalization has been shown to work better than batch normalization for the low batch size regime and has also been adopted by previous winners of the competition\cite{Myronenko18,Jiang19}. Unless specified otherwise, the number of group was set at 32. 

\subsubsection{Axial attention decoder}
\begin{figure}
    \centering
    \includegraphics[scale=0.26]{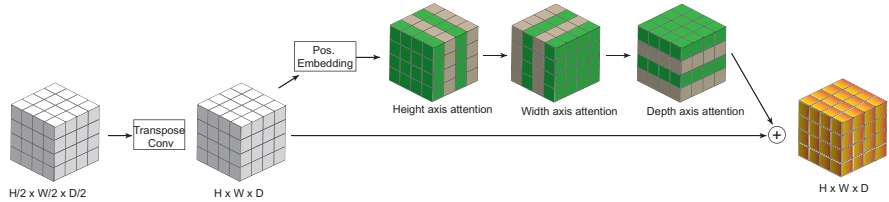}
    \caption{Upsampling with axial attention: axial attention is applied to each axis of the upsampled result from the transpose convolution. The output is added back to the original input, which is then concatenated with the features from the encoder path (not shown). }
    \label{fig:3}
\end{figure}

The final addition was using axial attention in the decoder. Self-attention or transformer \cite{Vaswani17} is a breakthrough idea that allows learning an adaptive attention of an input sequence based only on its self. Originally conceived and popularized in the NLP literatures \cite{Vaswani17,Devlin19,gpt3}, self-attention mechanism has slowly been adopted by the computer vision research community \cite{Dosovitskiy21}. One of the main obstacles when trying to apply self-attention to vision problems is that the computational complexity of the attention mechanism scales quadratically with the input size, rendering it impossible to fit or train the network in a standard workstation setup. This is even more of a problem when dealing with 3D data with the extra dimension. Axial attention \cite{Ho19,Wang20} has recently been proposed as an efficient solution when applying attention to multi-dimensional data. By applying self-attention to each axis of the input independently, the computation only scales linearly with image size, making it possible to integrate attention mechanism even with 3D data. We applied axial attention to the decoder of the network by running it on the output of the transposed convolution upsampling and then summing them. Fig. \ref{fig:3} showed an illustration of the axial attention decoder block. Even with more efficient attention, we found that it was not possible to apply the method to the highest resolution features ($128\times128\times128$) and opted for only the four lower resolutions. The number of attention heads and dimension of each head were doubled for each decrease in resolution, starting from 4 and 16 (at $64\times64\times64$ resolution), respectively.
\subsection{Training}
We followed the training methodology of nnU-Net for all networks. Each network was trained with 5-fold cross validation. During training, data augmentation was applied on the fly to improve the generalization. Data augmentation consisted of random rotation and scaling, elastic deformation, additive brightness augmentation, and gamma scaling. The objective for optimization is the sum of the binary cross entropy loss and the dice loss, calculated at the final full resolution output as well as at the auxiliary outputs of lower resolution. The batch version of the dice loss was used instead of the sample dice loss, computing the loss treating the whole batch as one sample instead of averaging dices from each sample in the minibatch. The batch dice helps stabilize the training by reducing the errors from samples with few annotated samples \cite{Fabian20}. The networks were optimized with stochastic gradient descent with Nesterov momentum of 0.99. The initial learning rate was 0.01 and was decayed following a polynomial schedule
\begin{equation}
    lr = 0.01\times(1 - \frac{epoch}{1000})^{0.9}
\end{equation}
Each training run lasted 1000 epochs, with each epoch consisting of 250 minibatches. The dice score on the validation set of the current fold was used to monitor the training progress. All experiments were conducted with Pytorch 1.9 on NVIDIA RTX 3090 GPU with 24GB VRAM. The following models were developed:

- \textbf{BL}: baseline nnUNet, batch normalization with batch size of 5

- \textbf{BL + L}: baseline with large Unet, batch size of 2, train on all training samples

- \textbf{BL + GN}: baseline with group normalization, batch size of 2

- \textbf{BL + AA}: baseline with axial attention, batch normalization, batch size of 2

- \textbf{BL + L + GN}: nnUNet with larger Unet, group normalization, batch size of 2
\section{Results}
\subsection{Quantitative Results}
Table \ref{table:1} showed the dice scores for the 3 tumor sub-regions of the 5 models from the cross-validation. Even though the model \textbf{BL + L} is invalid for this comparison due to being trained on the whole training dataset, increasing the size of the U-Net yielded a minor improvement. Using group normalization instead of batch normalization did not improve the performance and even slightly harm the dice metric. It is also worth noting that using group normalization increase memory consumption modestly, which might cancel out the memory reduction when using smaller batch size. Using the axial attention encoder did not improve the performance even at higher computation. Combining a large U-Net and group normalization increases the performance slightly for the tumor core and the whole tumor, with a large increase in GPU memory usage.

\newcolumntype{P}[1]{>{\centering\arraybackslash}p{#1}}
\begin{table}
\centering
\begin{tabular}{l|P{1cm}P{1cm}P{1cm}|c}
\centering
Model      &   ET   &   TC   &   WT   & Average \\ \hline
BL         &88.37    & 92.06   & 93.78   & 91.40    \\
BL + L     &89.82    & 94.03   & 94.58   & 92.81    \\
BL + GN    &88.17    & 92.11   & 93.66   & 91.30    \\
BL + AA    &87.23    & 91.88   & 93.21   & 90.77    \\
BL + L + GN &88.23   & 92.35   & 93.83   & 91.47       
\end{tabular}
\caption{\label{table:1}Dice metrics of the networks on 5-fold cross validation. ET: enhancing tumor, TC: tumor core, WT: whole tumor.}
\end{table}

Table \ref{table:2} showed the Dice and 95\% Hausdorff distance (HD95) computed by the competition organizers and displayed in the public leaderboard. All predicted segmentation labels were ensembled from 5 folds, except for the \textbf{BL + L} configuration. The ensembled results were then post-processed by converting the enhancing tumor class into necrotic tumor if the number of ET voxels is less than 200. Result for the axial attention model were not processed in time for the paper so it was omitted from the table. Incorporating the proposed changes led to a minor yet consistent improvement across all metrics when compared to the baseline model.
\begin{table}
\centering
\begin{tabular}{l|P{1cm}P{1cm}P{1cm}|P{1.5cm}|P{1cm}P{1cm}P{1cm}|P{1.5cm}}
\multirow{2}{*}{Model} & \multicolumn{4}{c|}{Dice} & \multicolumn{4}{c}{HD95} \\ \cline{2-9} 
                       & ET  & TC  & WT  & Average & ET  & TC  & WT  & Average \\ \hline
BL                     &83.73     &87.45     &92.63     &87.94         &22.44     &10.56     &3.55     &12.18         \\
BL + L                 &83.28     &86.53     &92.51     &87.44         &22.50     &12.75     &3.71     &12.99         \\
BL + GN                &84.09     &87.85     &92.77     &88.24         &22.41     &9.20      &3.42     &11.68         \\
BL + L + GN            &84.51     &87.81     &92.75     &88.36         &20.73     &7.623     &3.47     &10.61        
\end{tabular}
\caption{\label{table:2}Dice and 95\% Hausdorff distance of the networks as computed by the online evaluation platform. ET: enhancing tumor, TC: tumor core, WT: whole tumor, HD95: 95\% Hausdorff distance.}
\end{table}

\subsection{Qualitative Results}
\begin{figure}[h]
    \centering
    \includegraphics[scale=0.6]{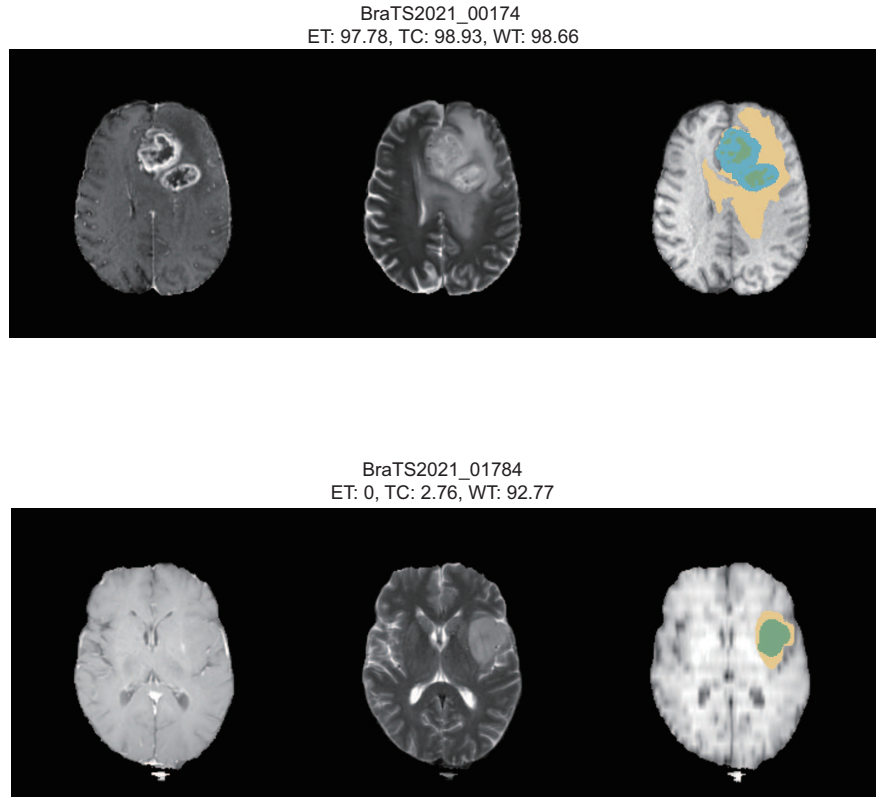}
    \caption{Representative good and bad predictions from the validation set. T1-GD, T2, and T1 with prediction from the network for two cases with good and bad metrics. Green: necrotic tumor core, blue: GD-enhancing tumor, yellow: edematous tissues.}
    \label{fig:4}
\end{figure}

Fig. \ref{fig:4} shows 2 representative examples of predictions from the \textbf{BL + L + GN} configuration. For the first case, the network successfully identified all the tumor sub-regions with high accuracy. This can potentially be attributed to the quality of the scans, with well defined contrasts for the tumor regions. For the second case, the network failed to segment the enhancing tumor and the tumor core while still performed decently for the whole extent of the tumor. Again, this can be explained by the quality of the MR scans: the T1 and T1-GD contrasts do not show clear delineation of the tumor and even have visible artifacts and blurring. This partially shows the importance of ensuring high quality data for the proper operation of the network. It also indicates that this data integrity factor should be considered when developing the networks, either through more diverse data acquisition or more meaningful data augmentation to cover these edge cases.

\section{Discussion}
In this short paper, we described our methodology for the BraTS 2021 competition. We extended the nn-UNet framework by using a larger network, replacing batch normalization with group normalization, and using axial attention decoder. These minor modifications slightly improve upon the baseline nn-UNet. As nn-UNet framework was extensively used for our method, we wanted to highlight the versatility, ease of use, and robustness of it. We were able to set up a very strong baseline without much experimentation. In fact, the pretrained networks that were developed on last year's data still performed well on this year's validation data, showing the generalizability of the method.

With 3D data, any modifications need to be carefully balanced with the availability of the GPU memory to ensure training can run without memory problem. The larger U-Net and axial attention decoder that we proposed can add significant memory footprint to the model even with small adjustment so it should be tuned carefully. Group normalization somewhat alleviates this issue by enabling the use of smaller batch size without incurring significant performance degradation.

It is important to inspect the failure cases to understand the behavior of the models. Most of the cases with inaccurate segmentation are similar to the case shown in Fig. \ref{fig:4}, with artifacts or quality issue in one of the contrasts. Other cases with bad Dice score are due to the post-processing method, which favors the removal of small enhancing tumor. The reason behind this was explained in detail in ref \cite{Fabian20}. For the short version, removing of small enhancing tumor can improve one's ranking on the leaderboard due to the dichotomy in the Dice and HD95 metrics when the ground truth segmentation does not have any enhancing tumor. We observed that using this post-processing method improves the Dice score slightly but worsen the HD95 score significantly for the enhancing tumor. We suspect this is due to the addition of several scans with small enhancing tumor, which render the post-processing harmful and worsen the HD95. We might need to consider more sophisticated post-processing method to address those cases.

\section{Acknowledgements}
We would like to acknowledge Fabian Isensee for his development of the nn-UNet framework and for sharing the models from last year competition.

%
%
%
%

\end{document}